\documentclass[sigconf]{acmart}

\usepackage{multirow}

\newcommand{\myet}{\emph{et~al.}}
\newcommand{\myeg}{\emph{e.g.}}
\newcommand{\myie}{\emph{i.e.}}

\AtBeginDocument{%
  }

\copyrightyear{2023} 
\acmYear{2023} 
\setcopyright{acmlicensed}\acmConference[MMAsia '23]{ACM Multimedia Asia
2023}{December 6--8, 2023}{Tainan, Taiwan}
\acmBooktitle{ACM Multimedia Asia 2023 (MMAsia '23), December 6--8, 2023,
Tainan, Taiwan}
\acmPrice{15.00}
\acmDOI{10.1145/3595916.3626397}
\acmISBN{979-8-4007-0205-1/23/12}

\begin{document}

\title{Cross-modal Consistency Learning with Fine-grained Fusion Network for Multimodal Fake News Detection}

\author{Jun Li}
\affiliation{%
  \institution{University of Electronic Science \\and Technology of China}
  \country{}
}
\email{y.lijun0321@hotmail.com}

\author{Yi Bin}
\affiliation{%
  \institution{National University of Singapore}
  \country{}
}
\email{yi.bin@hotmail.com}

\author{Jie Zou}
\affiliation{%
  \institution{University of Electronic Science \\and Technology of China}
  \country{}
}
\email{zoujie0806@gmail.com}

\author{Jiwei Wei}
\affiliation{%
  \institution{University of Electronic Science \\and Technology of China}
  \country{}
}
\email{mathematic6@gmail.com}

\author{Guoqing Wang}
\affiliation{%
  \institution{University of Electronic Science \\and Technology of China}
  \country{}
}
\email{gqwang0420@hotmail.com}

\author{Yang Yang}
\authornotemark[0]
\authornote{corresponding author}
\affiliation{%
  \institution{University of Electronic Science \\and Technology of China}
  \country{}
}
\email{yang.yang@uestc.edu.cn}


\begin{abstract}
Previous studies on multimodal fake news detection have observed the mismatch between text and images in the fake news and attempted to explore the consistency of multimodal news based on global features of different modalities.
However, they fail to investigate this relationship between fine-grained fragments in multimodal content.
To gain public trust, fake news often includes relevant parts in the text and the image, making such multimodal content appear consistent. 
Using global features may suppress potential inconsistencies in irrelevant parts.
Therefore, in this paper, we propose a novel Consistency-learning Fine-grained Fusion Network (CFFN) that separately explores the consistency and inconsistency from high-relevant and low-relevant word-region pairs.
Specifically, for a multimodal post, we divide word-region pairs into high-relevant and low-relevant parts based on their relevance scores.
For the high-relevant part, we follow the cross-modal attention mechanism to explore the consistency.
For low-relevant part, we calculate inconsistency scores to capture inconsistent points.
Finally, a selection module is used to choose the primary clue (consistency or inconsistency) for identifying the credibility of multimodal news.
Extensive experiments on two public datasets demonstrate that our CFFN substantially outperforms all the baselines.
\textit{Our code can be found at: \url{https://github.com/uestc-lj/CFFN/}}.
\end{abstract}

\ccsdesc[500]{Information systems~Social networks}

\keywords{Multimodal Fake News Detection, Fine-grained Fusion, Cross-modal Consistency Learning, Social Media}


\maketitle

\section{Introduction}
Due to its convenience and openness, online social media has gained popularity among people and has become the main platform for them to access and share information.
However, these advantages also lead social media to an ideal breeding ground for fake news. 
Previous studies~\cite{shu2017fake, zhou2020survey} have revealed that fake news always has negative impacts on democracy, justice, and public trust, \myeg, the willingness to receive the COVID-19 vaccine~\cite{montagni2021acceptance}, the trust on the result of the 2020 U.S. presidential election~\cite{calvillo2021individual}.
Therefore, the detection of fake news on online social media has emerged as an urgent necessity and has recently attracted much research attention~\cite{zhang2019multi, zhou2020similarity, qi2021improving, singhal2022leveraging, chen2022cross}.

\begin{figure}
	\centering		
	\includegraphics[width=0.8\linewidth]{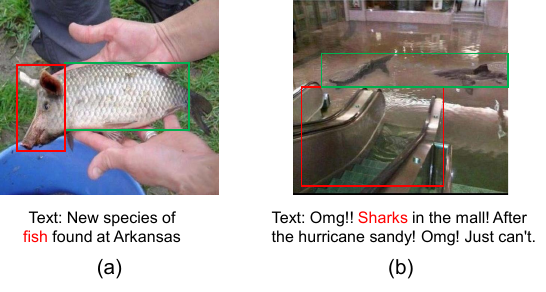}  
 \vspace{-5pt}
	\caption{Some examples of multimodal fake news. Compared with the high-relevant region in the green box, the red region is the low-relevant one, but provides the evidence (inconsistency) to identify the news as fake.} 
	\label{fig:motivation}
\end{figure}

Early studies~\cite{castillo2011information, yang2012automatic, kwon2013prominent, ruchansky2017csi,li2023focusing} mainly focused on textual content and further explored the social context, \myeg, propagation information and user responses, to detect fake news.
But these methods ignore the multimodal information of posts. 
Compared with the text-only posts, multimodal ones with attached images or videos could attract more attention and spread further and deeper~\cite{jin2016novel, jin2017multimodal}. 
Thus, considering both multimodal content and social context, att-RNN~\cite{jin2017multimodal} and MKEMN~\cite{zhang2019multi} are proposed.
However, they need to gather enough social context, which is difficult in the early stage of fake news dissemination, especially in the publishing phase~\cite{zhou2020similarity}.
Intuitively, for the multimodal posts, exploring the clues from the multimodal content is a practical direction. 

Recently, content-based approaches~\cite{zhou2020similarity, xue2021detecting, singhal2022leveraging} have observed the mismatch between text and images in fake news and proposed exploring the consistency between the textual and visual content in multimodal posts based on global features of different modalities.
However, these methods suffer from the following limitations:
(1) to obtain the global features, these methods leverage the pooling operation among all the fragments of a modality, which would suppress important information present in specific segments;
(2) these methods fail to align a word fragment with its relevant region fragments in the image and limit the detection performance without considering these fine-grained interactions.
To explore fine-grained interactions, several cross-modal attention-based studies~\cite{qian2021hierarchical, song2021multimodal} focus on high-relevant fragments from another modality and infer the clues from this relevant information.
However, such approaches overlook the significance of the less relevant and irrelevant information in multimodal news.
Typically, the text and the image in the real news are consistent and trustworthy, making them inherently relevant.
In the case of fake news, in order to gain the trust of the public, the creators undoubtedly strive to make multimodal content as relevant as possible.
As a result, the relevant information may provide limited evidence for making accurate decisions.
However, due to the falsified nature of the text or images in fake news, the less relevant or irrelevant portions of multimodal content may contain clues that reveal inconsistencies with our common knowledge.
As shown in Figure~\ref{fig:motivation}(a), compared with the body of the fish in the green box, the head of a pig (the red box) is less relevant to the focus word, ``fish'', in the text.
However, this less relevant word-region pair could serve as the inconsistency to confirm that this news is fake.
Similarly, in Figure~\ref{fig:motivation}(b), the escalator in the image is regarded as the inconsistent point of the word ``shark'', which reveals the mismatched living area of the shark in our real life. 

In this paper, we propose a novel Consistency-learning Fine-grained Fusion Network (CFFN) that extracts the consistency of real news from high-relevant word-region pairs and uncovers the inconsistency of fake news from low-relevant ones.
Specifically, for a text-image pair, we divide the word-region pairs into the consistent part and inconsistency-candidate part based on their relevance scores. 
The consistent part consists of high-relevant word-region pairs, while the inconsistency-candidate part contains word-region pairs with low relevance scores.
For the consistent part, we follow the cross-modal attention mechanism to fuse the relevant information for exploring consistency.
For the inconsistency-candidate part, we calculate the inconsistency scores for each word-region pair to capture the inconsistent points. 
Finally, a selection module is used to choose the primary clue (consistency or inconsistency) to identify the credibility of the multimodal news.
In addition, we construct a partition loss that encourages the model to select clues of the consistent part for recognizing real news and select clues of the inconsistency-candidate part for identifying fake news. 

The contributions of this paper are as follows:
\begin{itemize}
\item We propose to explore the consistency of real news from high-relevant word-region pairs and uncover the inconsistency of fake news from low-relevant word-region pairs.

\item For the multimodal news, we explore clues from both the consistent part and inconsistency-candidate part.
Additionally, we construct a partition loss that encourages the model to select clues of the consistent part for recognizing real news and clues of the inconsistency-candidate part for identifying fake news. 

\item Extensive experiments on two public datasets demonstrate that our proposed CFFN could achieve the best performance among all the baselines.
\end{itemize}

\section{Method}
\begin{figure*}
	\centering		
	\includegraphics[width=0.9\linewidth]{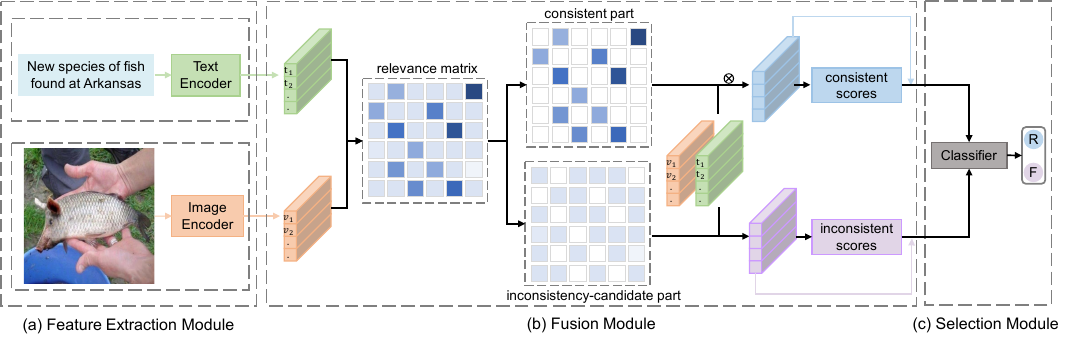}  
	\caption{Illustration of the proposed CFFN. Our model consists of three main components: feature extraction module, fusion module, and selection module. The fusion module splits all the word-region pairs into the consistent part and the inconsistency-candidate part based on the relevance scores. Then the consistency and the inconsistency are respectively explored from these two parts. Finally, the selection module captures the main clue (consistency or inconsistency) to identify the credibility of the multimodal news (R: real news; F: fake news).} 
	\label{fig:framework}
\end{figure*}

The overview of our proposed model, CFFN, is illustrated in Figure~\ref{fig:framework}. 
Specifically, CFFN consists of three main components: feature extraction module, fusion module, and selection module.
Our model aims to separate the word-region pairs into two parts based on the relevance score. We explore the consistency from the high-relevant part and the inconsistency from the low-relevant part.
In this section, we will elaborate each component of our CFFN.

\subsection{Feature Extraction Module}
The multimodal news consists of a textual modality and a visual modality.
In this work, we focus on the text and the image, which are denoted as $W$ and $R$.
To extract better fine-grained fragment features, we leverage the pre-trained model as our feature encoder, which has the great power to understand the contextual information in a modality.

\subsubsection{Text Encoder}
Given a text, the pre-trained BERT~\cite{devlin2018bert} could explore the bidirectional context of each word and obtain its representation, so we utilize it as our text encoder to extract the fine-grained word representation, which denoted as:
\begin{equation}
	E_t = BERT(\{w_1, w_2, \cdots, w_N\}),
\end{equation}
where $N$ is the length of the text and $E_t = \{e_t^1, e_t^2, \cdot, e_t^N\} \in \mathbb{R}^{N \times d_t}$ is the word embedding matrix.
Similar to~\cite{qu2021dynamic}, 1-D convolution kernels with different sizes, \myie, \{1, 2, 3\}, are used to obtain the phrase-level information for each word.
Afterwards, the set of phrase-level information is projected into the word features with a full-connected layer, \myie, $T=\{t_1, t_2, \cdots, t_N\} \in \mathbb{R}^{N \times d}$.

\subsubsection{Image Encoder}
Recently, Swin Transformer~\cite{liu2021swin} has shown great potential on multiple vision tasks, such as object detection and cross-modal retrieval~\cite{xu2023multi,bin2023unifying}.
Compared with other Transformer-based vision models, Swin Transformer could construct hierarchical feature maps with shifted windows, so we utilize it as our image encoder to extract region features, which are denoted as:
\begin{equation}
	E_v = Swin\text{-}T(R),
\end{equation}
where $E_v = \{e_v^1, e_v^2, \cdots, e_v^M\} \in \mathbb{R}^{M \times d_v}$ is the region features matrix, $M$ is the number of regions in the image, and $d_v$ is the dimension of the extracted region features.
Afterwards, we project $E_v$ into the same space of word features with a full-connected layer, \myie, $V=\{v_1, v_2, \cdots, v_M\} \in \mathbb{R}^{M \times d}$.

\subsection{Fusion Module}
Attention mechanism is widely used in natural language processing~\cite{vaswani2017attention, sun2020understanding}, computer vision~\cite{chen2016attention,bin2021entity, wang2017residual,yang2018video}, and cross-modal tasks, \myeg, cross-modal retrieval~\cite{liu2019focus, lee2018stacked,li2023your} and visual question answering~\cite{peng2019cra, wang2020multimodal,peng2021progressive}. 
Similar to the cross-modal retrieval, several fake news detection methods~\cite{qian2021hierarchical, song2021multimodal, wu2021multimodal} leverage the cross-modal attention to select the relevant complementary information from another modality for a given fragment.
For instance, for a word $w_i$, they use it as a query to calculate the attention scores with all the regions in the image and the higher score means highly relevance:
\begin{equation}
    \left\{ \quad
    \begin{aligned}
        s_j &= \sigma(t_i \, v_j^T), \\
        \hat{t_i} &= \sum^M_{j=1}s_j \, v_j,
    \end{aligned}
    \right.
\end{equation}
where $t_i$ is the word features of $w_i$, $s_j$ is the attention score between the word feature $t_i$ and the region feature $v_j$, $M$ is the number of the regions in the image, $\hat{t_i}$ is the complementary information of $w_i$ in the visual modality, and $\sigma$ is the softmax activation.

These methods focus on the high-relevant information from another modality and use it to infer the consistency of the multimodal content.
However, they overlook the significance of the low-relevant parts for fake news detection.
In these approaches, the low-relevant parts receive lower attention scores during the fusion process, which may result in missing crucial clues, \myeg, potential inconsistencies, in fake news.
Typically, to gain public trust, fake news often incorporates high-relevant parts in the multimodal content, creating an appearance of coherence and consistency to mislead the audience.
Therefore, this high-relevant information may provide limited evidence for identifying fake news.
In contrast, the low-relevant parts may contain crucial clues that reveal inconsistencies with common knowledge.
These subtle inconsistencies in the low-relevant fragments can be significant indicators for detecting fake news, such as the word, ``Sharks'', and the region of the escalator in Figure~\ref{fig:motivation}(b).
Hence, in this work, we attempt to explore the consistency of real news from the high-relevant part and the inconsistency of fake news from the low-relevant part.

Inspired by~\cite{zhou2020similarity,xue2021detecting}, we leverage the cosine similarity to assess the relevance scores between the fragments in different modalities.
For word features $T=\{t_1, t_2, \cdots, t_N\} \in \mathbb{R}^{N \times d}$ and region features $V=\{v_1, v_2, \cdots, v_M\} \in \mathbb{R}^{M \times d}$, the relevance score is denoted as:
\begin{equation}
	s_{ij} = \frac{t_i \, v_j^T}{||t_i|| \, ||v_j||}, \quad i \in [1, N], j \in [1, M],
\end{equation}
where $s_{ij} \in [-1, 1]$ is the relevance score of the word feature $t_i$ and region feature $v_j$.

To separate out the low-relevant fragments, we leverage a threshold $\lambda \in [0, 1)$ to split the relevance matrix $S=\{s_{11}, s_{12}, \cdots, s_{NM}\} \in \mathbb{R}^{N \times M}$ into two parts:
\begin{equation}
    \left\{ \quad
	\begin{aligned}
		&S_m = \{{s_{ij} > \lambda}\}, \\
		&S_c = \{{s_{ij} \leq \lambda}\}, 
	\end{aligned}
	\right.
\end{equation}
where $S_m$ denotes the consistent part where the word-region pairs are highly relevant, while $S_c$ denotes the inconsistency-candidate part where the word-region pairs are weakly relevant or irrelevant. 

For word-region pairs in the consistent part, we refer to the attention mechanism and aggregate them with their relevant scores:
\begin{equation}
    \hat{t_i} = \sum^M_{j=1}\sigma (s_{ij}) \, v_j + t_i, \quad \{i, j\} \in S_m,
\end{equation}
where $\hat{t_i}$ is the complementary information in the visual modality.
As the image contains background information that contains no relevant words in the text, in this paper, we mainly use the word as the query to select the relevant regions. 

As for the word-region pairs in the inconsistency-candidate part, since low-relevant fragments are provided, we leverage the element-wise addition to obtain the representations $c_{ij}$:
\begin{equation}
    c_{ij} = t_i \oplus v_j, \quad \{i, j\} \in S_c,
\end{equation}
where $\oplus$ denotes the element-wise addition.

To explore the clues of the consistent part and inconsistency-candidate part, we attempt to calculate consistent scores and inconsistent scores respectively, and then aggregate the features in the corresponding part with such scores.
Concretely, we map $c_{ij}$ into a scalar score with an MLP layer to produce the inconsistent score:
\begin{equation}
    s_{ij}^c = \textbf{sigmoid}(MLP(c_{ij})),
\end{equation}
where MLP has the hidden dimension $d_m$ and the hyperbolic tangent activation function.
Similarly, we transform $\hat{t_i}$ into a consistent score $s_{i}^m$.
Finally, the aggregate process is denoted as:
\begin{equation}
    \left\{ \quad
	\begin{aligned}
		&z_m = \sum^N_{i=1} s_{i}^m \, \hat{t_i}, \quad, i \in S_m, \\
		&z_c = \sum^N_{i=1}\sum^M_{j=1} s_{ij}^c \, c_{ij}, \quad, \{i, j\} \in S_c, 
	\end{aligned}
	\right.
\end{equation}
where $z_m \in \mathbb{R}^{1 \times d}$ and $z_c \in \mathbb{R}^{1 \times d}$ denote the representations of the consistent part and the inconsistency-candidate part respectively.

\subsection{Selection Module}
Taking the inconsistency-candidate representation $z_c$ and the consistent part representation $z_m$ as the input, the selection module assigns a weight to each representation, in order to evaluate which part provides the primary clue for the final decision:
\begin{equation}
    \left\{ \quad
	\begin{aligned}
        &w_m = (W_z \, z_m + b_z), \\
        &w_c = (W_z \, z_c + b_z), \\ 
		&w_{mc} = \sigma (w_m \circ w_c), 
	\end{aligned}
	\right.
\end{equation}
where $W_z \in \mathbb{R}^{d \times 1}$ and $b_z \in \mathbb{R}^{1}$ are the training parameters, $\circ$ denotes the concatenation operation, and $w_{mc} \in \mathbb{R}^{1 \times 2}$.
The multimodal content representation $z$ is denoted by the weight aggregation of $z_c$ and $z_m$:
\begin{equation}
    z = w_{mc} [z_m \circ z_c]^T,
\end{equation}
where $[z_m \circ z_c] \in \mathbb{R}^{d \times 2}$.
Afterwards, the classifier is used to predict the credibility $\hat{y}$ of the multimodal content by transforming its representation $z$ into two classes with a two-layer perceptron:
\begin{equation}
    \hat{y} = W_{f_2} \, (ReLU(W_{f_1} \, z + b_{f_1})) \, + b_{f_2}, 
\end{equation}
where $W_{f_2} \in \mathbb{R}^{d_f \times 2}, b_{f_2} \in \mathbb{R}^{2}, W_{f_1} \in \mathbb{R}^{d_f \times d}, b_{f_1} \in \mathbb{R}^{d_f}$ are the training parameters.
We denote $y$ as the ground-truth label of the multimodal content and train the modal with the detection loss:
\begin{equation}
    \mathcal{L}_d = -[y \ log \hat{y} + (1-y) \ log(1-\hat{y})].
\end{equation} 

Additionally, in this paper, we attempt to explore the consistency of real news from the highly relevant part and the inconsistency of fake news from the weakly relevant part.
Therefore, to encourage the model to select clues from the right part, we construct the partition label $y_p$ and design a partition loss function as:
\begin{equation}
    \mathcal{L}_p = ||y_p - w_{mc}||^2_2,
\end{equation} 
note that $w_{mc}$ is a two-dimensional vector with the first dimension representing the weight of the consistent part and the second dimension representing the weight of the inconsistency-candidate part. 
Therefore, the label $y_p = [1, 0]$ is for real news to enlarge the weight of the consistent part, while $y_p = [0, 1]$ is for fake news to expand the weight of the inconsistency-candidate part. 

Finally, our total loss function is designed with the linear combination of $\mathcal{L}_d$ and $\mathcal{L}_p$:
\begin{equation}
    \mathcal{L} = \mathcal{L}_d + \beta \mathcal{L}_p,
\end{equation} 
where $\beta \in (0.0, 1.0] $ is a trade-off parameter.

\section{EXPERIMENTS}
\begin{table*}
	\centering
	\caption{The performance of different models on two datasets.}
	\label{tab:result}
	\begin{tabular}{c ccccccc c ccccccc}
        \hline
        & \multicolumn{7}{c}{\textbf{Weibo}} && \multicolumn{7}{c} {\textbf{Twitter}} \\  
        \hline
        \multirow{2}{*}{\textbf{Method}} & \multirow{2}{*}{\textbf{Acc}} & \multicolumn{3}{c}{\textbf{Fake News}} & \multicolumn{3}{c}{\textbf{Real News}} && \multirow{2}{*}{\textbf{Acc}} & \multicolumn{3}{c}{\textbf{Fake News}} & \multicolumn{3}{c}{\textbf{Real News}}\\
        \cmidrule(r){3-5} \cmidrule(r){6-8} \cmidrule(r){11-13} \cmidrule(r){14-16}
        && \textbf{Pre} & \textbf{Rec} & \textbf{F1} & \textbf{Pre} & \textbf{Rec} & \textbf{F1} &&& \textbf{Pre} & \textbf{Rec} & \textbf{F1} & \textbf{Pre} & \textbf{Rec} & \textbf{F1} \\
        \hline
        CNN & 0.740 & 0.736 & 0.756 & 0.744 & 0.747 & 0.723 &  0.735 && 0.549 & 0.508 & 0.597 & 0.549 & 0.598 & 0.509 &  0.550\\
        VS & 0.726 & 0.732 & 0.712 & 0.722 & 0.720 & 0.740 & 0.730 && 0.617 & 0.635 & 0.644 & 0.639 & 0.639 & 0.630 & 0.634\\
        \hline
        att\_RNN & 0.772 & 0.854 & 0.656 &  0.742 & 0.720 & 0.889 & 0.795 && 0.664 & 0.749 & 0.615 &  0.676 & 0.589 & 0.728 & 0.651\\
        EANN & 0.782 & 0.827 & 0.697 & 0.756 & 0.752 & 0.863 & 0.804 && 0.648 & 0.810 & 0.498 & 0.617 & 0.584 & 0.759 & 0.660\\
        MVAE & 0.824 & 0.854 & 0.769 & 0.809 & 0.802 & 0.875 & 0.837 && 0.745 & 0.801 & 0.719 & 0.758 & 0.689 & 0.777 & 0.730\\
        SpotFake & 0.869 & 0.877 & 0.859 & 0.868 & 0.861 & 0.879 & 0.870 && 0.771 & 0.784 & 0.744 & 0.764 & 0.769 & 0.807 & 0.787\\
        MKEMN & 0.814 & 0.823 & 0.799 & 0.812 & 0.723 & 0.819 & 0.798 && 0.715 & 0.814 & 0.756 & 0.708 & 0.634 & 0.774 & 0.660\\
        SAFE & 0.816 & 0.818 & 0.815 & 0.817 & 0.816 & 0.818 & 0.817 && 0.766 & 0.777 & 0.795 & 0.786 & 0.752 & 0.731 & 0.742 \\
        MCNN & 0.823 & 0.858 & 0.801 & 0.828 & 0.787 & 0.848 & 0.816 && 0.784 & 0.778 & 0.781 & 0.779 & 0.790 & 0.787 & 0.788 \\
        HMCAN & 0.885 & \textbf{0.920} & 0.845 & 0.881 & 0.856 & \textbf{0.926} & 0.890 && 0.897 & \textbf{0.971} & 0.801 & 0.878 & 0.853 & \textbf{0.979} & 0.912\\
        CAFE & 0.840 & 0.855 & 0.830 & 0.842 & 0.825 & 0.851 & 0.837 && 0.806 & 0.807 & 0.799 & 0.803 & 0.805 & 0.813 & 0.809\\       
        \hline
        CFFN(Res) & 0.891 & 0.913 & 0.869 & 0.890 & 0.871 & 0.914 & 0.892 && 0.906 & 0.853 & 0.946 & 0.897 & 0.954 & 0.875 & 0.913\\
        CFFN(VGG) & 0.893 & 0.914 & 0.876 & \textbf{0.893} & 0.874 & 0.915 & 0.894 && 0.908 & 0.867 & 0.930 & 0.897 & 0.943 & 0.890 & 0.916\\
        CFFN & \textbf{0.901} & 0.913 & \textbf{0.889} & 0.889 & \textbf{0.888} & 0.913 & \textbf{0.900} && \textbf{0.923} & 0.872 & \textbf{0.965} & \textbf{0.916} & \textbf{0.971} & 0.891 & \textbf{0.929}\\
        \hline
    \end{tabular}
\end{table*}

\subsection{ Experimental Setup}
\subsubsection{Datasets}
We train and evaluate our model on two public datasets: Twitter~\cite{boididou2018detection} and Weibo~\cite{jin2017multimodal}.
The Twitter dataset was released for Verifying Multimedia Use task, a part of MediaEval~\cite{2016Verifying}.
Each tweet in this dataset consists of textual content, visual content (image/video), and corresponding social context. 
Following previous work~\cite{jin2017multimodal, khattar2019mvae}, we remove the tweets with videos in the dataset.
As the Twitter dataset has already been split into a development set and a test set, we use the development set for training and another for testing.
As for Weibo dataset, Jin~\myet~\cite{jin2017multimodal} collected the fake news from the official rumor debunking system of Weibo and the real news from Xinhua News Agency.
We follow preprocessing steps and the same data split of~\cite{jin2017multimodal, khattar2019mvae, qian2021hierarchical} in our experiments.

\subsubsection{Baseline Methods}
We compare the proposed model with two groups of baselines: unimodal models and multimodal models, which are listed as follows:

\textbf{Unimodal models:}
(1) \textbf{CNN}~\cite{yu2017convolutional} employs a two-layer CNN to learn the feature representations for misinformation identification;
(2) \textbf{VS}~\cite{jin2016novel} focuses on learning different image distribution patterns of real and fake news by several visual and statistical features.

\textbf{Multimodal models:}
(1) \textbf{att\_RNN}~\cite{jin2017multimodal} proposes Recurrent Neural Network with an attention mechanism to fuse the multimodal features including textual, visual, and social context features. 
We remove the social context features for a fair comparison in our experiments;
(2) \textbf{EANN}~\cite{wang2018eann} derives event-invariant features by the event discrimination task and utilizes these features to detect fake news on newly arrived events;
(3) \textbf{MVAE}~\cite{khattar2019mvae} uses a variational autoencoder to learn the shared feature vector of the multimodal input and detect fake news with such feature vector;
(4) \textbf{SpotFake}~\cite{singhal2019spotfake} leverages the pre-trained BERT as the textual encoder and pre-trained VGG19 as the visual encoder to extract the global textual and visual representations, and concatenates them as the multimodal features for fake news detection;
(5) \textbf{MKEMN}~\cite{zhang2019multi} constructs a memory bank with the textual, visual, and knowledge information of the existing events and could retrieve the event-invariant features from this bank for the unseen events;
(6) \textbf{SAFE}~\cite{zhou2020similarity} transforms the visual content to a caption and detects fake news by exploring the similarity relationship between the text content and the caption;
(7) \textbf{HMCAN}~\cite{qian2021hierarchical} proposes a hierarchical multimodal contextual attention network to fuse the output of different layers of pre-trained BERT and visual features with the cross-attention mechanism;
(8) \textbf{MCNN}~\cite{xue2021detecting} extracts textual features, visual semantic features, and visual tampering features for exploring the consistency of multimodal content;
(9) \textbf{CAFE}~\cite{chen2022cross} regards the cross-modal ambiguity as a gate to adaptively aggregate unimodal features or capture cross-modal correlations.

\subsubsection{Implementation Details}
In our experiments, the textual encoder inherits huggingface’s implementation\footnote{\url{https://github.com/huggingface/transformers}} and the visual encoder is the tiny Swin Transformer~\cite{liu2021swin} model called Swin-T.
We set the text length to at most 200 words for the Weibo dataset and 50 words for the Twitter dataset and use the special token [PAD] in BERT embedding to fill the text less than the setting length.
For the text content, we leverage 1-D CNN with a set of kernel sizes, \myie, {1, 2, 3} to transform the output of BERT from 768 to 256 (the embedding space).
For the visual content, we first resize the image into 224x224x3 and use Swin-T to obtain the region embedding $M \times d_v$, \myie, $49 \times 768$.
Then a full-connected layer is applied to transform the region embedding into the same space with the text embedding, \myie, 256.
In the fusion module, we respectively set $\lambda$ as 0.0 and 0.1 for Weibo and Twitter.
As for the selection module, the hidden dimensions $d_m$ and $d_f$ of MLP are set to 128 and 64 respectively.
Finally, the model is optimized by Adam optimizer with the learning rate of 0.001 and the weight decay of $10^{-4}$.
The batch size for both datasets is 128.

\subsection{Overall Performance}
Table~\ref{tab:result} demonstrates the performance comparison of our proposed CFFN and other unimodal and multimodal baselines on the Twitter dataset and Weibo dataset.  
We have the following observations: 

First, for the unimodal methods, VS and CNN explore clues from the images and the text, respectively.
We find that VS is better than CNN on Twitter, while CNN is better than VS on Weibo. 
This implies that, for different datasets, visual content and textual content may play different roles.
It is essential to consider the content of both modalities for fake news detection.

Second, due to the excellent power of the pre-trained models, methods such as SpotFake and HMCAN, with the pre-trained language model and pre-trained visual model, outperforms other methods. 
Besides, the fine-grained method, HMCAN, outperforms SAFE, CAFE, and SpotFake, which employ the global features of each modality after the pooling operation.
These observations indicate that the pre-trained models and the fine-grained interaction between different modalities are both important for improving detection performance. 

Finally, CFFN leverages the pre-trained models, BERT and Swin Transformer, to fuse different modalities in a fine-grained way and reaches the best accuracy, 0.901 for Weibo and 0.923 for Twitter.
By splitting the word-region pairs into consistent and inconsistency-candidate parts, CFFN could explore the consistency and inconsistency from these parts and further make the right decision.
Additionally, as the pre-trained VGG~\cite{simonyan2014very} and ResNet~\cite{he2016deep} are widely used for extracting the visual feature on fake news detection, we also leverage them as the visual encoder, named CFFN(Res) and CFFN(VGG).
These two variants both outperform the baselines, which indicates the advantages of our method. 
However, they are both worse than CFFN (with Swin-T), proving that a better visual encoder could substantially improve the performance.
   
\subsection{Ablation Study}
To validate the effectiveness of the components in CFFN, we design some ablation experiments on both datasets.

\subsubsection{Effectiveness of Different Components in CFFN}
\begin{table}
	\centering
	\caption{The effectiveness of components in CFFN.}
	\label{tab:ablation_component}
	\begin{tabular}{c cccc}
        \hline
        \multirow{2}{*}{\textbf{Method}} & \multirow{2}{*}{\textbf{Dataset}} & \multirow{2}{*}{\textbf{Acc}} & \multicolumn{2}{c}{\textbf{F1}} \\
        \cline{4-5}
        &&& \textbf{Fake} & \textbf{Real}\\
        \hline
        \multirow{2}{*}{\textbf{w/o consistent part}} & Weibo & 0.884 & 0.885 & 0.883\\
        & Twitter & 0.920 & 0.912 & 0.926 \\        
        \hline
        \multirow{2}{*}{\textbf{w/o inconsistent part}} & Weibo & 0.883 & 0.884 & 0.881\\
        & Twitter & 0.906 & 0.896 & 0.915\\        
        \hline
        \multirow{2}{*}{\textbf{w/o partition loss $\mathcal{L}_p$}} & Weibo & 0.882 & 0.882 & 0.882\\
        & Twitter & 0.913 & 0.904 & 0.920 \\        
        \hline
        \multirow{2}{*}{\textbf{w/o separation}} & Weibo & 0.876 & 0.878 & 0.875\\
        & Twitter & 0.902 & 0.888 & 0.913\\        
        \hline
        \multirow{2}{*}{\textbf{CFFN}} & Weibo & 0.901 & 0.889 & 0.900\\
        & Twitter & 0.923 & 0.916 & 0.929\\        
        \hline
    \end{tabular}
\end{table}

As shown in Table~\ref{tab:ablation_component}, we design four variants of CFFN: (1) \textbf{w/o consistent part:} a variant only uses the inconsistency-candidate part to explore the inconsistency; 
(2) \textbf{w/o inconsistent part:} a variant only captures the consistency in the consistent part for fake news detection; 
(3) \textbf{w/o partition loss $\mathcal{L}_p$:} a variant could adaptively infer clues from the consistent part and inconsistency-candidate part; 
(4) \textbf{w/o separation:} a variant directly leverages cross-modal attention to fuse relevant regions as the complementary information for each word in the text content.

Among all the variants, \textbf{w/o separation} yields the poorest performance , which indicates the effectiveness of separating the word-region pairs to explore the consistency and inconsistency in fake news detection.
Additionally, we observe that:
(1) Both \textbf{w/o consistent part} and \textbf{w/o inconsistent part} yields poorer performance than \textbf{CFFN}.
This proves that exploring the consistency from the consistent part and the inconsistency from the inconsistency-candidate part are both essential. 
(2) Compared with \textbf{CFFN}, the performance of \textbf{w/o partition loss $\mathcal{L}_p$} drops about 1\% and 1.9\%, proving that the effectiveness of assigning the model to mine the consistency for real news and infer the inconsistency for fake news.

\subsubsection{Impact of the Hyperparameters}
\begin{figure}
	\centering		
	\includegraphics[width=\linewidth]{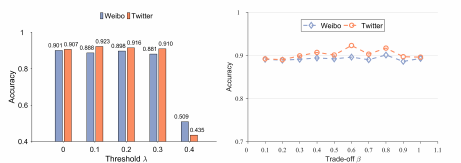}  
	\caption{The impact of the trade-off $\beta$ and the threshold $\lambda$ on the performance for Weibo and Twitter datasets.} 
	\label{fig:tradeoff}
\end{figure}

We have two main hyperparameters, \myie the trade-off parameter $\beta$ and the threshold $\lambda$, to train our CFFN.  
Specifically, $\beta$ controls the supervision of assigning the consistent part for real news and the inconsistency-candidate part for fake news.
A smaller value of $\beta$ represents more freedom for CFFN to select clues from both the consistent part and inconsistency-candidate part, while a larger value of $\beta$ would restrict the model to infer clues from the corresponding part.
As shown in Figure~\ref{fig:tradeoff}, the results reveal that our model could reach the best accuracy when $\beta=0.6$ and $\beta=0.8$ for the Twitter dataset and Weibo dataset respectively.

As for the threshold $\lambda$, it is the critical parameter to separate the consistent part and inconsistency-candidate part.
To contain the low-relevant word-region pairs in the inconsistency-candidate part, we design the range of $\lambda$ is $[0.0, 1.0)$.
During the comparison experiments, we observe that the performance of the CFFN drops significantly when $\lambda \geq 0.4$ , which is identifying all the news as fake news.
This results from the cosine similarity of a word and a region in these two datasets is smaller than 0.4.
When $\lambda \geq 0.4$, all the word-region pairs belong to the inconsistency-candidate part.
The partition loss restricts the model to explore the clues from the consistent part for real news. However, there are no word-region pairs in the consistent part, so the model regards all the news as fake news. 
Therefore, we set $\lambda \in [0.0, 0.3]$. 
The results in Figure~\ref{fig:tradeoff} show that our model could reach the best accuracy when $\lambda=0.1$ and $\lambda=0.0$ for the Twitter dataset and Weibo dataset respectively.

\begin{figure}
	\centering		
	\includegraphics[width=\linewidth]{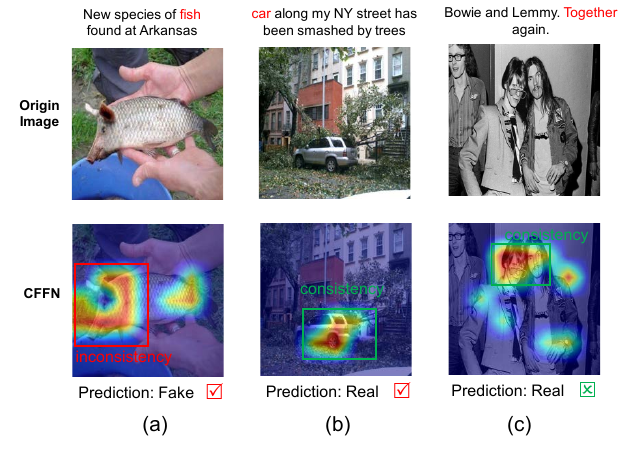}  
	\caption{The qualitative evaluation of our CFFN. The focus of the text is marked in red, and its corresponding consistency or inconsistency in the image is also marked with a green box or a red box. Among three examples, (a) and (b) are the cases the model predicts correctly, while (c) is the failure one.} 
	\label{fig:case}
\end{figure}

\subsection{Qualitative Evaluation}
In CFFN, the inconsistency and consistency are essential to identify fake news and real news respectively.
We quantify the focus of our CFFN on multimodal contents of different news in Figure~\ref{fig:case}.
Among the three examples, for fake news in Figure~\ref{fig:case} (a), our model focuses on the inconsistency points, \myie, head of the pig in the image and the ``fish'' in the description text, which are in the inconsistency-candidate part.
In contrast, for the word ``car'' in the text of Figure~\ref{fig:case} (b), CFFN captures the consistent region in the image for identifying the real news and this word-region pair belongs to the consistent part.
The failure case in (c) reveals the limitation of our model.
Lacking the external knowledge, CFFN cannot recognize the persons in the picture and only captures the consistency of the word ``Together'' and the region where two persons are standing together, thereby making the wrong decision, \myie, real news.

\section{CONCLUSIONS}
In this paper, we proposed a novel Consistency-learning Fine-grained Fusion Network (CFFN) to explore both the consistency and inconsistency from the fine-grained fragments, \myie, word-region pairs, in the multimodal content.
Unlike the cross-modal attention mechanism, which focused on the high-relevant information from another modality and overlooked the significance of the low-relevant parts, our CFFN attempted to explore the consistency and inconsistency respectively from high-relevant and low-relevant word-region pairs.
Finally, the primary clue (consistency or inconsistency) was used to identify the credibility of the multimodal news.
Besides, the hyperparameter, $\lambda$, is important in our model to construct different parts, but it is fixed for all the samples in a dataset.
In the future, we will further explore to learn the threshold considering the different distributions of word-region pairs in consistency and inconsistency parts.  

\begin{acks}
This work was partially supported by the National Natural Science Foundation of China under grant 62220106008, U20B2063, and 62102070. This work was also partially supported by Sichuan Science and Technology Program under grant 2023NSFSC1392.
\end{acks}

\clearpage
\bibliographystyle{ACM-Reference-Format}
\bibliography{myref}

\end{document}